\documentclass[sigplan,10pt,screen]{acmart}
\renewcommand\footnotetextcopyrightpermission[1]{}
\AtBeginDocument{%
  }

\usepackage{tikz}
\usepackage{enumitem}
\usetikzlibrary{arrows.meta, positioning, decorations.pathmorphing, calc, patterns}
\usepackage{amsmath}
\usepackage{subcaption}
\usepackage{multirow}
\usepackage{pifont}
\usepackage{listings}
\usepackage[linesnumbered, noend, ruled, vlined]{algorithm2e}
\usepackage{xcolor}
\usepackage{subcaption}
\usepackage{multirow}
\usepackage{array}
\usepackage{graphicx}
\usepackage{xcolor}
\usepackage{color}
\definecolor{ZpfGreen}{RGB}{0,100,0}
\definecolor{ZpfRed}{RGB}{255,0,102}

\usepackage{filecontents}

\usepackage{color}
\definecolor{ZpfGreen}{RGB}{0,100,0}
\definecolor{ZpfRed}{RGB}{255,0,102}

\newcommand{\stitlestart}[1]{\noindent{\bf #1\/}}

\newcommand{\stitle}[1]{\vspace*{0.3em}\noindent{\bf #1\/}}

\newcommand{\name}{{\textsf{OmniCast}}}

\newcommand{\squishlist}{
	\begin{list}{$\bullet$}
		{ \setlength{\itemsep}{1pt}
			\setlength{\parsep}{1pt}
			\setlength{\topsep}{2.5pt}
			\setlength{\partopsep}{0.5pt}
			\setlength{\leftmargin}{1em}
			\setlength{\labelwidth}{1em}
			\setlength{\labelsep}{0.6em}
		}
	}
	\newcommand{\squishend}{
	\end{list}
}

\expandafter\def\expandafter\normalsize\expandafter{%
    \normalsize%
    \setlength\abovedisplayskip{2pt}%
    \setlength\belowdisplayskip{4pt}%
    \setlength\abovedisplayshortskip{-4pt}%
    \setlength\belowdisplayshortskip{2pt}%
}

\definecolor{XiangyuRed}{RGB}{200,30,30}

  {\par\medskip\begingroup\color{XiangyuRed}\noindent\textbf{[xiangyu]}\enspace\ignorespaces}%
  {\par\endgroup\medskip}

\begin{document}

\title{OmniCast: Interaction-Aware Serving for Real-Time Omni-Modal LLMs}

\author{Xiangyu Zhi}
\authornote{These authors contributed equally to this work.}
\affiliation{%
  \institution{The Chinese University of Hong Kong}
  \country{}
}
\email{xyzhi24@cse.cuhk.edu.hk}

\author{Peiqi Yin}
\authornotemark[1]
\affiliation{%
  \institution{The Chinese University of Hong Kong}
  \country{}
}
\email{pqyin22@cse.cuhk.edu.hk}

\author{Sheng Guan}
\affiliation{%
  \institution{The Chinese University of Hong Kong}
  \country{}}
\email{codelformat@gmail.com}

\author{Chenguang Zheng }
\affiliation{%
  \institution{The Chinese University of Hong Kong}
  \country{}}
\email{chenguangzheng@link.cuhk.edu.hk}

\author{James Cheng}
\affiliation{%
  \institution{The Chinese University of Hong Kong}
  \country{}}
\email{jcheng@cse.cuhk.edu.hk}

\author{Xiao Yan}
\affiliation{%
  \institution{Wuhan University}
  \country{}}
\email{yanxiaosunny@whu.edu.cn}

%


\begin{abstract}
Realtime omni-modal LMs support speech-centric conversations where users stream inputs, hear generated audio, and interrupt freely. Existing Omni-LM serving systems still rely on throughput-oriented LLM scheduling and LRU KV offloading. These policies ignore audio playback and multi-turn reuse: they may generate tokens far beyond what users hear, wasting work after barge-in, and evict KV state needed in the next turn. OmniCast is an interaction-aware serving system for realtime Omni-LM interaction. It exposes playback progress, speech activity, and barge-in events to the serving pipeline. The scheduler prioritizes first-audio and near-underrun sessions while limiting generation beyond the playback frontier. The KV manager uses next-use-aware eviction and preloads likely-needed KV during user speech to hide reload latency. On vLLM-Omni, OmniCast improves realtime serving across two Omni-LMs and mixed workloads. It lowers P90 audio TTFP by $1.55\times$ on average and up to $2.21\times$, while improving completed-request throughput by $1.15\times$ on average and up to $1.56\times$, and moves most KV reload work off the next-turn critical path.
\end{abstract}




\maketitle

\makeatletter
\fancyhead[LE]{}
\fancyhead[RO]{}
\makeatother

\section{Introduction}
\label{sec:intro}

\emph{Omni-modal large models} (Omni-LMs)~\cite{Qwen3-Omni, Ming-Flash-Omni, LongCat-Flash-Omni, mimo-omni, gemini-omni} (also known as \emph{any-to-any} models) extend large language models (LLMs)~\cite{gpt4, gemini, qwen3} to process and produce multi-modal contents (e.g., text, image, video, and audio). 
Representative Omni-LMs include the Qwen-Omni series~\cite{Qwen3-Omni, Qwen2.5-Omni, Qwen3.5-omni}, Ming-Omni~\cite{Ming-Flash-Omni, Ming-Omni}, and Nemotron VoiceChat~\cite{nemotron-voice}. These models typically adopts a multi-stage pipeline instead of a monolithic decoder: the upstream stages perform multimodal understanding, mapping heterogeneous inputs into shared token or embedding representations, while the downstream stages synthesize the user-facing outputs.
For instance, Qwen3-Omni~\cite{Qwen3-Omni} utilizes modality-specific \emph{encoders} to handle the inputs, a language backbone (\emph{thinker}) for reasoning and response planning, and speech synthesis components (\emph{talker} and \emph{vocoder}) to produce audible user replies.

\begin{figure}[!t]
  \centering
  \includegraphics[width=\linewidth]{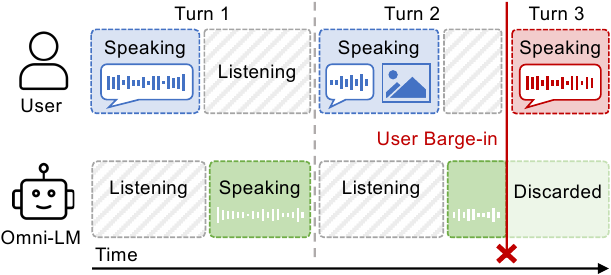}
  \caption{Interactive Omni-LM serving with multiple turns.}
  \label{fig:omni-realtime-workload}
  \vspace{-2mm}
\end{figure}

As shown in Figure~\ref{fig:omni-realtime-workload}, Omni-LM serving is usually \textit{interactive} with multiple turns (called a \textit{session}). In particular, the user provides streaming speech, image, and video inputs to the Omni-LM and may barge in (i.e., interrupt) while the model is still generating responses. The user-model interaction can take multiple turns, and the model takes all  previous turns as contexts to process the current turn. 
Such an interactive paradigm is adopted by applications like hands-free assistants~\cite{Omnigaia, wen2026evostreaming}, real-time translation~\cite{Openomni}, customer support~\cite{didas2026multi}, and accessibility services~\cite{MATE}.
Commercial offerings include Google's Gemini Live~\cite{GeminiLive}, OpenAI's Realtime API~\cite{openai-live}, and ByteDance's Doubao Realtime Voice~\cite {doubao-realtime-voice}. 


The community currently adapts high-throughput LLM serving systems to run the multiple stages (i.e., model components) of Omni-LMs in a \emph{stage-oriented} manner.
Representative stacks include vLLM-Omni~\cite{vllm-omni} and SGLang-Omni~\cite{sglang-omni}, which extend the runtime of vLLM~\cite{vllm} and SGLang~\cite{sglang}, respectively.
In these systems, different model stages (e.g, encoders, thinker, and talker) run on separate inference engines with independent schedulers, batching policies, GPU placement, and possibly different parallelization strategies, and an orchestrator routes intermediate results between the stages. 
These systems optimize the individual stages for metrics like time-to-first-token (TTFT) and time-between-tokens (TBT). Lacking a global view at the session level with multiple interactive stages, these systems suffer from two critical problems that hinder efficiency.

\squishlist
\item \textit{Excessive generation.} LLM serving engines optimize metrics like TTFT and TBT for individual stages, which do not necessarily lead to better user experience during Omni-LM interactions. Specifically, for live audio reply, users care the most about audio time-to-first-packet (audio TTFP) and whether the playback is smooth.
Once the audio starts, users are largely satisfied as long as  the playback has no glitches, and further reducing the TBT does not improve user experience.
However, existing Omni-LM serving systems conduct continuous token generation at each LLM stage without considering the playback of the audio stream, and many tokens could be generated ahead of the playback progress. 
If users barge in during the playback, the generated but not yet played tokens will be wasted.

\item \textit{Passive cache management.} For each stage of Omni-LMs, a session needs to keep all KV caches of the previous turns for the current turn, stressing limited GPU HBM when serving multiple sessions concurrently. Audio and video inputs make the matter worse by producing large KV caches. LLM serving systems can offload the KV caches to CPU DRAM~\cite{vllm-kv-offload} but eviction is handled reactively via LRU without considering realtime interaction semantics. For instance, a session that is quiet during user playback may appear cold and be evicted, although its KV cache will likely be needed for the next interaction turn. Moreover, evicted KV cache is restored only when the current turn already starts, and thus  reloading time  boosts TTFP.
\squishend

To tackle the two problems above, we build \name{}, an interaction-aware serving system for Omni-LMs. Compared with existing solutions that process the stages of Omni-LMs separately, \name{} adopts a global view at the session level to make holistic request batching decisions via \textit{urgency based scheduling} and handle KV cache swapping with \textit{proactive cache management}.
In particular, to avoid excessive generation while ensuring user experience, \name{} adds an interaction layer in vLLM-Omni to  track real-time session states, including playback progress, speech activity, and barge-in events.
Using these signals, \name{} schedules the inference requests for the next batch based on the playback progresses of their sessions. Sessions that have not produced the first output token and sessions whose playback buffer is close to under-run are prioritized to reduce audio TTFP and avoid glitches, respectively. In contrast, sessions whose generated tokens are  far ahead of the  playback progress are delayed to avoid wasting computation if the users barge in.

Different from LRU that conducts eviction passively according to historical cache accesses, \name{} proactively utilizes  interaction patterns to infer future cache accesses. Specifically, \name{} keeps the KV caches in GPU HBM for sessions whose audio playback is coming to an end since their KV caches will soon be used  for the next interaction. Meanwhile, sessions that have just started a long playback are offloaded to CPU memory because they are likely to remain inactive for a while. Moreover, when users begin speaking or barge in, \name{} preloads the CPU resident KV caches to GPU HBM as these sessions will become active for generation shortly. 
Such a design overlaps DRAM-to-HBM transfer with user's speaking time and moves cache reloading latency off the next-turn audio TTFP path.

We implement \name{} atop vLLM-Omni and evaluate its performance using multiple Omni-LMs models and realtime workloads.
Experimental results show that \name{} improves the maximum sustainable throughput by up to $55\%$ over vLLM-Omni while maintaining the same audio TTFP. 
In the presence of random user barge-in, our urgency based scheduling reduces P90 audio TTFP by over $50\%$ and cuts the calculated-but-unheard tokens by $75\%$ on average.
Meanwhile, proactive cache management reduces the peak LLM-stage KV residency by about $26\%$ and moves most KV reload work off the next-turn critical path.

To summarize, we make the following contributions:

\squishlist
  \item We analyze the problems of existing systems when serving interactive Omni-LMs with continuous playback, possible barge-in, and multi-turn context reuse.
  \item We propose and build \name{}, an interaction-aware serving system for Omni-LMs, adopting a holistic session-level view over individual stages to enhance efficiency.
  \item We design urgency based scheduling to ensure user experience while avoiding wasting computation and proactive cache management to handle KV cache eviction and hide DRAM-HBM cache loading latency.
\squishend

\section{Background and Motivation}
\label{sec:background}

\subsection{Omni-modal  Models}
\label{sec:omni-model}

\begin{figure}[t]
  \centering
  \includegraphics[width=\linewidth]{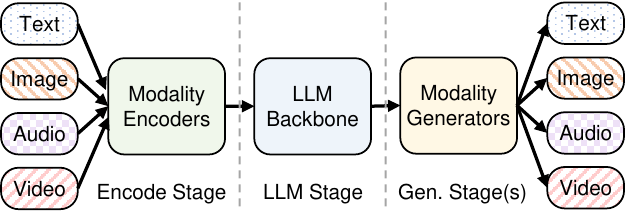}
  \caption{A common architecture of Omni-LMs.}
  \label{fig:omni}
\end{figure}

Text-only LLMs excel at language understanding and generation, but are limited to textual inputs and outputs.
\emph{Omni-modal large models} (Omni-LMs)~\cite{Qwen3-Omni, Ming-Flash-Omni, LongCat-Flash-Omni, mimo-omni, gemini-omni}, also referred to as \emph{any-to-any} multimodal models, extend this capability by jointly \emph{understanding} and \emph{generation} content across text, images, video, and audio within unified architectures.
This unification supports flexible cross-modal reasoning and multimodal output beyond separate understanding and generation pipelines, including spoken responses as well as diffusion-based image or video generation~\cite{gemini-omni, hunyuanimage}.

The emergence of any-to-any modeling leads to substantially more complex structures than conventional LLMs.
As shown in Figure~\ref{fig:omni}, modern Omni-LMs commonly organize inference into an \emph{encoding stage}, an autoregressive (AR) \emph{LLM stage}, and one or more \emph{generation stages}.
The encoding stage maps multimodal inputs into embedding representations consumed by the LLM backbone.
The backbone performs semantic understanding and response planning, then drives modality generators such as speech talkers and vocoders, or diffusion transformers (DiT)~\cite{DiT} for visual synthesis.

Recent Omni-LMs instantiate this template with different choices of encoders, language backbones, and output generators.
Qwen3-Omni~\cite{Qwen3-Omni} is a representative speech-oriented design.
It accepts text, image, video, and audio inputs through dedicated encoders, uses a \emph{thinker} LLM for understanding and response planning, and routes hidden states to a \emph{talker} that generates speech tokens followed by a lightweight \emph{vocoder} for waveform reconstruction.
Many other Omni-LMs follow the same multi-stage organization while swapping in different generators for their target modalities, such as the Ming-Omni series~\cite{Ming-Omni, Ming-Flash-Omni} and LongCat-Flash-Omni~\cite{LongCat-Flash-Omni} for spoken interaction, Baichuan-Omni-1.5~\cite{li2025baichuan} for controllable speech output, or AR-plus-DiT pipelines for image and video outputs~\cite{hunyuanimage, GLMImage2026}.
The concrete modules differ across models, but the overall architecture remains modular, with encoders for ingestion, a shared backbone for reasoning, and specialized decoders for each output modality.

To reduce response latency, recent models use different methods of incremental execution, but they differ in how much cross-stage overlap they expose.
Some models only stream input or output tokens, while others propose \emph{asynchronous chunking} across adjacent stages~\cite{Qwen3-Omni, nemotron-voice}.
Take Qwen3-Omni as an example. The thinker can pass partial hidden-state chunks to the talker before thinker decoding fully completes, and the talker can pass speech-token chunks to the vocoder before the full sequence is generated.
This chunked execution lowers audio TTFP and improves user-perceived responsiveness in spoken interaction.

\begin{figure}[t]
  \centering
      \setlength{\abovecaptionskip}{0.05cm}
  \includegraphics[width=0.9\linewidth]{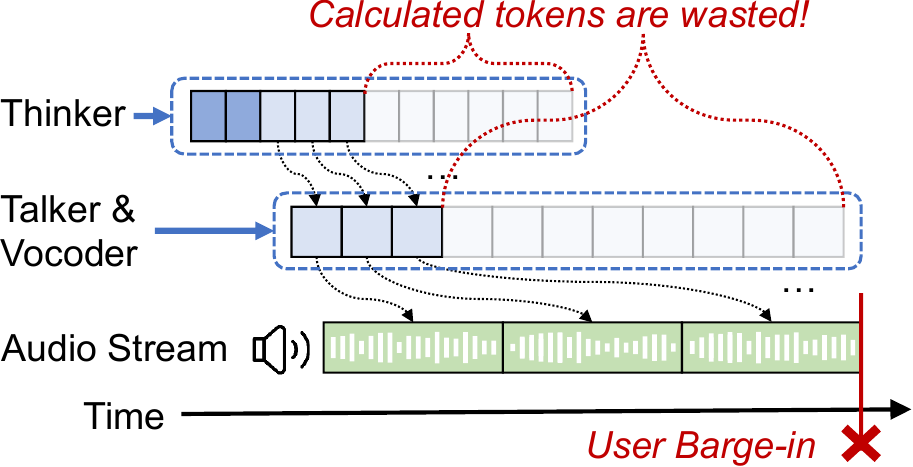}
  \caption{Interaction-unaware generation ahead of playback.}
  \Description{Timeline of audio playback, thinker generation, and talker/vocoder generation until user barge-in wastes calculated tokens.}
  \label{fig:bargein}
\end{figure}

\subsection{Serving Systems for Omni-LMs}
\label{sec:omni-serving}

To deploy Omni-LMs in production, recent systems extend high-throughput LLM runtimes to multi-stage pipelines~\cite{vllm-omni,sglang-omni}.
vLLM-Omni~\cite{vllm-omni} builds on vLLM~\cite{vllm} and introduces a \emph{fully disaggregated} serving stack.
Users decompose an Omni-LM into interconnected stages (e.g., encoders, thinker, talker, vocoder, or DiT modules); each stage runs as an independent engine with its own scheduler and KV cache manager.
An orchestrator drives request progress across the stages, while inter-stage connectors route intermediate tensors and control metadata between engines.
As stages have different compute and memory profiles, each can adopt an independent parallelization strategy, such as data parallelism or tensor parallelism, without constraining the rest of the pipeline.
SGLang-Omni~\cite{sglang-omni} pursues a similar stage-oriented design on top of SGLang~\cite{sglang}.
Both systems treat Omni-LM inference as coordinated execution across multiple stages rather than a single monolithic forward pass, improving job completion time and resource utilization for diverse any-to-any workloads.
They primarily provide the serving support for general inference scenarios with multimodal generation, such as text-to-speech, audio chat, and image generation.

\subsection{Serving Challenges}
\label{sec:challenges}

The interactive workload in Section~\ref{sec:intro} creates two serving problems. The server keeps generating far ahead of playback, and multi-turn KV is evicted and reloaded at the wrong time.
We next examine both problems in turn.

\stitle{Interaction-unaware generation ahead of playback.}
During a spoken reply, the client plays audio at a fixed rate, while the thinker and talker continue decoding and synthesizing speech ahead of what has been heard.
Stage-local schedulers still favor faster token production, so they keep extending this generation lead even once playback already has enough buffered audio.
If the user barges in, the unplayed portion of that lead is wasted GPU work, as Figure~\ref{fig:bargein} illustrates.

\begin{figure}[t]
  \centering
    \setlength{\abovecaptionskip}{0.05cm}
  \includegraphics[width=\linewidth]{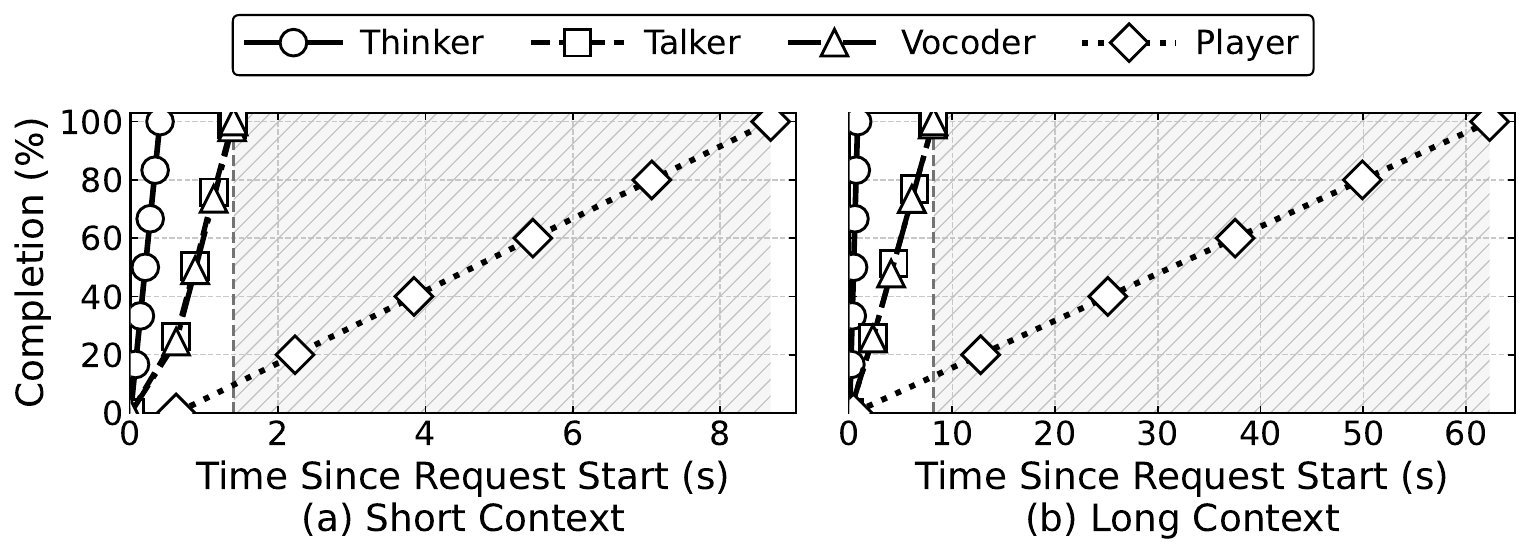}
  \caption{Generation and playback completion over time.}
  \Description{Line charts with time on the x-axis and completion percentage on the y-axis, showing thinker, talker, vocoder, and player progress for short and long contexts.}
  \label{fig:gen-playback-completion}
  \vspace{-3mm}
\end{figure}

Figure~\ref{fig:gen-playback-completion} quantifies this execution-time mismatch on Qwen3-Omni.
The thinker, talker, and vocoder reach full completion quickly, while client playback progresses much more slowly.
Server-side stages therefore finish far ahead of what the user has actually heard, leaving a long interval in which the system keeps generating tokens that may never be played.
A barge-in during this interval discards a large portion of already-computed work, and the waste grows when a longer context makes execution finish further ahead of playback.

\stitle{Interaction-unaware multi-turn KV management.}
A multi-turn session must keep prior-turn KV at each stage so the next request can reuse the full context, and audio or video inputs make this resident state grow quickly against limited HBM.
Existing engines spill idle session KV to host DRAM~\cite{vllm-kv-offload} and use \textit{least recently used (LRU)} to decide which blocks remain in GPU memory.
LRU reflects recent access rather than upcoming turn timing, so a session can be evicted while the user is still listening and then pay a reload penalty when the next turn begins.

\begin{figure}[t]
  \centering
  \includegraphics[width=\linewidth]{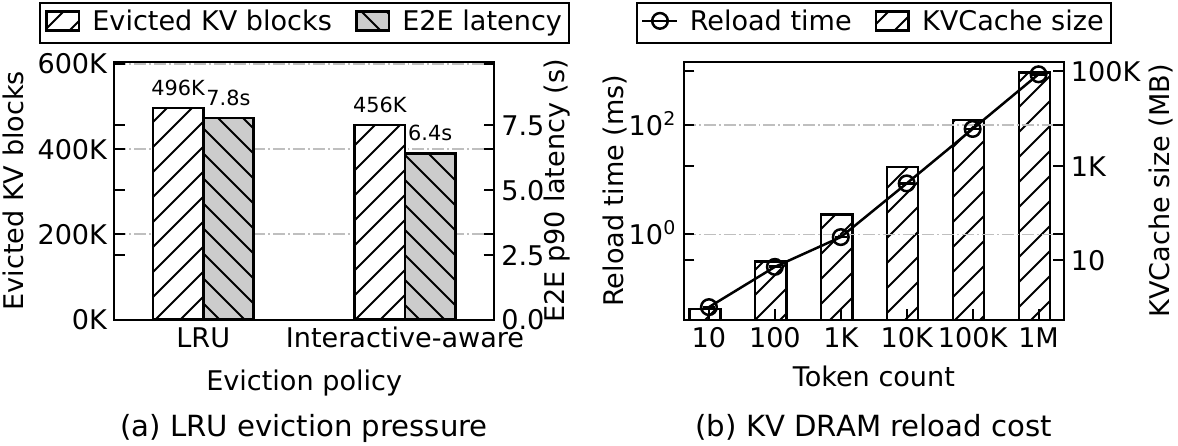}
  \caption{Interaction-unaware multi-turn KV management. (a)~LRU eviction under load increases evicted KV blocks and tail latency. (b)~Reloading offloaded KV from host DRAM back to GPU HBM incurs latency that grows with KV size.}
  \Description{Two-panel chart showing that LRU eviction under load increases evicted KV blocks and end-to-end tail latency, and that DRAM KV reload time grows with token count and KV cache size.}
  \label{fig:kv-scaling}
\end{figure}

Figure~\ref{fig:kv-scaling} shows that this mismatch becomes more costly as load increases.
First, a session that is temporarily quiet during playback may appear cold and be evicted even though its KV will likely be needed on the next turn, causing hot requests to be evicted and increasing both reload volume and E2E tail latency under concurrent load (Figure~\ref{fig:kv-scaling}(a)).
Second, reload timing is not coordinated with interaction (Figure~\ref{fig:kv-scaling}(b)).
When KV loading from host DRAM begins only after the user has already started the next turn, the transfer sits on the critical path to LLM prefill and directly inflates audio TTFP; this cost grows with the amount of offloaded KV that must be brought back to GPU HBM.
Because loading is not interaction-aware, the system misses chances to preload KV during client-side idle intervals, so reactive reload at turn start makes TTFP much worse than necessary.

\section{OmniCast Architecture}
\label{sec:arch}

\begin{figure}[t]
  \centering
    \setlength{\abovecaptionskip}{0.05cm}
  \includegraphics[width=0.95\linewidth]{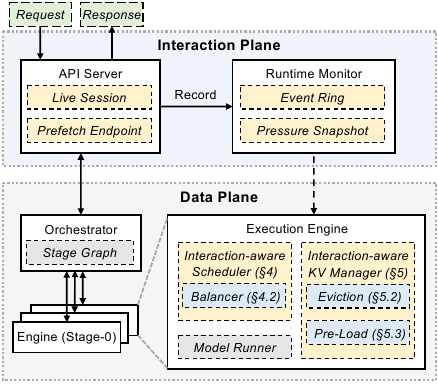}
  \caption{System architecture of \name{}.}
  \Description{Architecture diagram with an interaction plane containing the API server, live session interface, prefetch endpoint, and runtime monitor with an event ring and pressure snapshot; and a data plane containing the orchestrator, stage graph, and execution engines. Each execution engine contains an interaction-aware scheduler, an interaction-aware KV manager, and a model runner.}
  \label{fig:liveserve-arch}
\end{figure}

We propose \name{}, an interaction-aware serving system for realtime Omni-LM interaction.
As shown in Figure~\ref{fig:liveserve-arch}, \name{} keeps the stage-oriented execution model of vLLM-Omni, but separates the runtime into an \emph{interaction plane} and a \emph{data plane}.
The data plane preserves the orchestrator and stage-local engines, while the interaction plane tracks live session signals such as streaming arrivals, playback progress, and barge-in events.
By exposing these signals to schedulers and KV managers, \name{} allows engine-level control decisions to react to user interaction without changing the model pipeline.
Below, we describe how this separation is realized through the session-facing interaction layer, interaction-aware execution engines, and stage-oriented orchestration.

\squishlist
\item \textit{Session-facing interaction layer.}
The API server is the entry point for live multimodal sessions, forwarding streaming inputs to the orchestrator and returning generated audio to clients.
It also exposes the prefetch endpoint for history preheat and current-turn prefill, labeling this preparatory work separately from latency-critical decoding.
Alongside the API server, a lightweight runtime monitor tracks how each session is progressing at the client, including whether playback is advancing normally or has been interrupted.
The monitor turns these client-side signals into a compact runtime view that engine policies can read without directly coupling to the session protocol.

\item \textit{Interaction-aware execution engines.}
Each model stage runs as an execution engine, such as the thinker, talker, and vocoder engines in a speech-oriented Omni-LM.
Inside each engine, the scheduler manages the stage-local request queue and forms batches, while the KV manager keeps the cached conversation state used by that stage.
\name{} feeds session state from the runtime monitor into these two components, so scheduling and KV-residency decisions can be aware of client-side interaction.
This keeps interaction-aware control inside the engine where the corresponding execution and memory decisions are made.

\item \textit{Stage-oriented orchestration.}
\name{} adopts vLLM-Omni's orchestrator and stage graph to connect these engines into an Omni pipeline.
The orchestrator routes requests, intermediate data, and generated outputs across stages, while each stage run with its own engine and GPU allocation.
Thus, \name{} preserves the existing disaggregated execution substrate and focuses its changes on the interaction-aware control inside the engines.
\squishend

\stitle{Barge-in handling.}
During playback, the client reports how much generated audio has been consumed, while VAD detects whether the user starts a new utterance.
When a barge-in arrives, the orchestrator notifies the engines serving the current response; the engines abort in-flight computation, discard tokens beyond the playback point, and clear temporary stage state.
In parallel, the runtime monitor records the interruption in the session state seen by later scheduling and KV-management decisions.
This keeps abort and cleanup on the execution path, while letting interaction-aware policies observe the interruption through the monitor.

\section{Interaction-aware Request Scheduling}
\label{sec:scheduling}

\begin{figure}[t]
  \centering
  \includegraphics[width=0.95\linewidth]{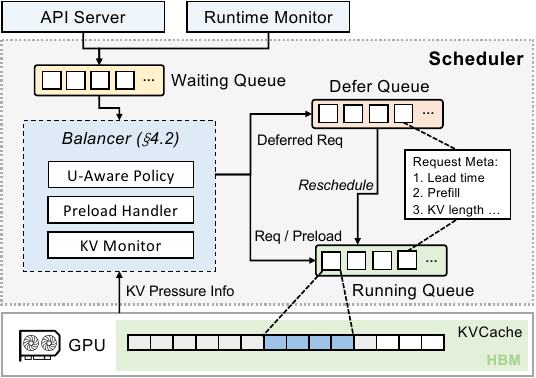}
  \caption{Overview of the interaction-aware scheduler.}
  \Description{Scheduler overview showing monitor signals feeding an urgency classifier, U0/U1/U2 request classes, U2 utility ordering, and feasible batch selection for an execution engine.}
  \label{fig:scheduler-overview}
\end{figure}

Modern LLM serving systems commonly use first-come-first-serve (FCFS) request scheduling inside each execution engine~\cite{vllm,sglang}.
Requests wait until the engine has enough round budget to admit them into the running set, after which continuous batching repeatedly schedules active requests for decode steps.
This design keeps GPU batches large and improves throughput.
It also reduces time-between-tokens (TBT), which matters for text generation because users observe the response token by token.

Realtime Omni-LM interaction changes what the scheduler should protect.
Before playback starts, asynchronous chunking improves audio TTFP only if downstream stages are scheduled soon after their first chunks become ready.
After playback starts, the primary goal is to keep audio playback smooth.
Generating farther ahead of the client does not further improve the user's immediate experience and can waste work if the user interrupts.
FCFS misses both cases, since it may continue serving well-buffered sessions while other sessions are still waiting for first audio or are close to playback underrun.

\begin{figure}[t]
  \centering
  \includegraphics[width=1\linewidth]{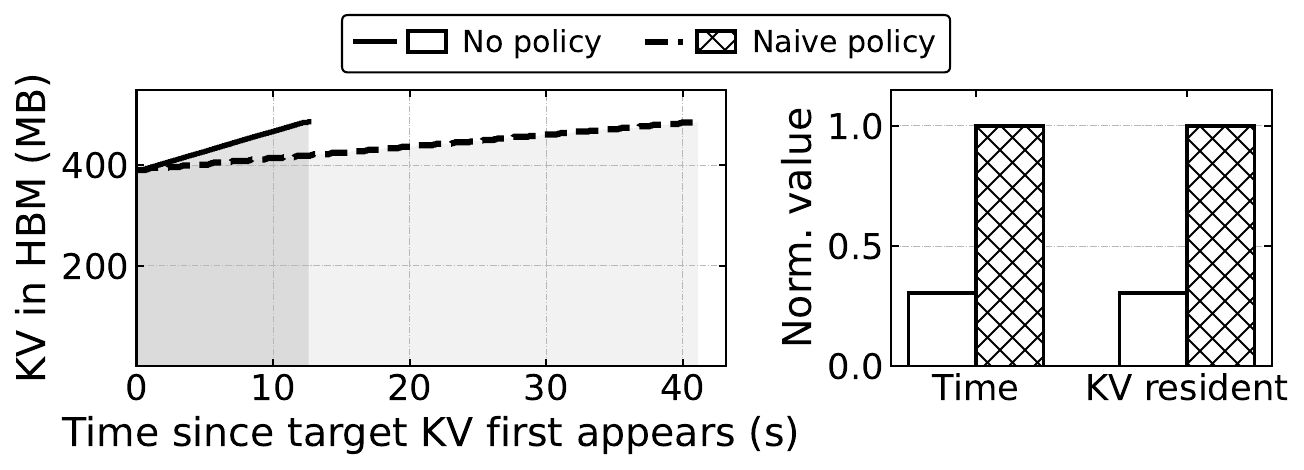}
  \caption{Motivating KV-pressure-aware scheduling with a long multi-turn dialogue. The left panel tracks the GPU KV-cache residency of one long-context request over time, while the right panel reports normalized residency duration and average resident KV footprint.}
  \Description{Two-panel plot comparing a no-policy baseline and a naive U0-only policy. The left panel shows the target request's GPU KV-cache footprint over time, and the right panel shows normalized residency time and average resident KV footprint.}
  \label{fig:active-request-kv-residency}
\end{figure}

\subsection{Urgency-Aware Scheduling}
\label{sec:scheduling-objective}

\name{} replaces FCFS-style ordering with an interaction-aware policy inside each execution engine.
At each scheduling round, the scheduler reads session state from the runtime monitor and chooses a resource-feasible subset from its ready set $\mathcal{R}_s$.
Serving a request too late can delay its first playable audio or let playback drain, while serving a well-buffered request too aggressively may produce work that the user never hears.
As shown in Figure~\ref{fig:scheduler-overview}, \name{} first separates requests by realtime urgency and then ranks requests within the same urgency class.

The scheduler assigns request $i$ at stage $s$ to an urgency class $\Gamma_i^s\in\{0,1,2\}$, where smaller values indicate higher urgency.
The main signal is the \textit{stage-aware playback buffer} $\mathcal{P}_i^s$, which estimates how much work stage $s$ has produced beyond what downstream consumers currently need.
A positive buffer means the stage has produced more units than the downstream consumer has consumed or is ready to consume.
For audio-generation stages, this buffer corresponds to playable audio waiting at the client, computed from generated audio and client playback progress.
For upstream stages, it is estimated from downstream work queues, such as talker-ready or pending talker work.
Let $\mathcal{P}_{\mathrm{safe}}^s$ be the minimum safe buffer used to identify playback risk.

\squishlist
\item \textit{U0: playback urgency.}
Requests that have started playback but have a small playback buffer ($\mathcal{P}_i^s\le\mathcal{P}_{\mathrm{safe}}^s$) are most urgent ($\Gamma_i^s=0$), as delaying them may cause audible stalls.
\name{} sorts them by playback buffer in ascending order, serving the session closest to underrun first.
\item \textit{U1: first-audio urgency.}
Requests that have not produced their first playable audio packet are next ($\Gamma_i^s=1$), because they are still on the audio TTFP critical path.
\name{} sorts them by ready age $\mathcal{A}_i$, preserving FCFS-style aging within this class.
\item \textit{U2: efficiency scheduling.}
Requests that have already produced first audio and have enough playback buffer ($\Gamma_i^s=2$) fall into this class.
\name{} orders them with a utility that balances KV-pressure relief against excessive generation ahead of playback.
\squishend

\stitle{U2 efficiency policy.}
\name{} orders U2 requests with a lightweight utility that combines KV-pressure relief and barge-in exposure.
Figure~\ref{fig:active-request-kv-residency} illustrates why the KV term is needed in a long multi-turn dialogue.
We compare a no-policy baseline with an urgency-only scheduler that protects U0 playback-risk requests but ignores the KV state of non-urgent resident requests.
The urgency-only scheduler can leave the target request resident much longer after it leaves the playback-critical path, because it is repeatedly delayed behind U0 work while its large context remains allocated.
This prolonged residency consumes allocatable KV blocks and increases HBM contention under memory pressure.
The barge-in term addresses the other U2 goal by penalizing requests that already have enough playback headroom but keep generating far ahead of the client.
For request $i$ at stage $s$, the utility is
\begin{equation}
\mathcal{U}_i^s =
\beta_s \mathcal{U}_{\mathrm{kv},i}^s
-
\alpha_s \mathcal{C}_{\mathrm{barge},i}^s,
\label{eq:u2-utility}
\end{equation}
where $\alpha_s$ and $\beta_s$ tune the relative weight of barge-in exposure and KV-pressure relief for stage $s$.
\name{} sets their ratio offline: for each stage, we sweep candidate $\alpha_s/\beta_s$ values on a mock workload and choose the setting that gives the best tradeoff between discarded work and KV-pressure stalls.

The barge-in exposure cost penalizes large playback buffers, since additional calculated-but-unheard work may be discarded if the user interrupts.
\name{} computes
\begin{equation}
\mathcal{C}_{\mathrm{barge},i}^s =
\frac{\max(0,\mathcal{P}_i^s-\mathcal{P}_{\mathrm{safe}}^s)}
{\mathcal{P}_{\mathrm{safe}}^s}.
\label{eq:barge-exposure}
\end{equation}
This cost grows when the request exceeds the safe buffer, so requests with normal playback headroom are not penalized.

The KV-pressure benefit favors requests that already occupy substantial KV when the stage KV pool is close to full.
Let $\mathcal{K}_i^s$ denote the GPU KV cache occupied by request $i$, and let $\mathcal{R}_{s,\mathrm{occ}}$ denote the ratio between occupied GPU KV and allocatable GPU KV at this stage.
\name{} computes
\begin{equation}
\mathcal{U}_{\mathrm{kv},i}^s =
\mathcal{K}_i^s
\cdot
\mathcal{R}_{s,\mathrm{occ}}.
\label{eq:kv-pressure-benefit}
\end{equation}
This benefit grows when a long resident request is holding KV in a crowded stage.
\name{} uses $\mathcal{U}_i^s$ only as a per-round ordering heuristic for U2 requests after realtime urgency has been handled.

\subsection{Scheduling Procedure}
\label{sec:scheduling-policy}
Each execution engine runs the policy at the beginning of a scheduling round, as shown in Algorithm~\ref{alg:interactive-scheduling}.
The ready set $\mathcal{R}_s$ contains requests that can make progress at stage $s$ if they are admitted into the next batch.
Before selecting the batch, the scheduler refreshes the per-request interaction state and estimates the playback buffer $\mathcal{P}_i^s$ for each ready request.
It then assigns each request an urgency class $\Gamma_i^s\in\{0,1,2\}$ according to the rules above.
Only U2 requests need the efficiency score $\mathcal{U}_i^s$, which is computed after classification from the current playback buffer and KV pressure.

The scheduler forms the next batch by sorting each urgency class, joining the three lists in U0-U1-U2 order, and scanning the candidates.
For each candidate, if admitting it would exceed the remaining round budgets $\mathcal{M}_s$, namely the token budget and available KV blocks, admission stops.
Otherwise, \name{} adds the request to $\mathcal{B}_s$ and reduces $\mathcal{M}_s$ by the token and KV capacity that the request consumes.
Requests that are not selected remain in engine state and are reconsidered in the next scheduling round.

\begin{algorithm}[t]
\small
\DontPrintSemicolon
\KwIn{ready set $\mathcal{R}_s$, arrival ages $\mathcal{A}$, round budgets $\mathcal{M}_s$}
\KwOut{next batch $\mathcal{B}_s$}
\BlankLine
\SetKwProg{Fn}{Function}{:}{}
\Fn{\textsc{Schedule}$(\mathcal{R}_s,\mathcal{A},\mathcal{M}_s)$}{
  $\mathcal{B}_s\leftarrow\emptyset$\;
  \ForEach{$i\in\mathcal{R}_s$}{
    $\mathcal{P}_i^s\leftarrow\textsc{EstimateBuffer}(i)$\;
    $\Gamma_i^s\leftarrow\textsc{Classify}(i,\mathcal{P}_i^s)$\;
    \If{$\Gamma_i^s=2$}{
      $\mathcal{U}_i^s\leftarrow\textsc{ComputeUtility}(i,\mathcal{P}_i^s)$\;
    }
  }
  $\mathcal{C}_0\leftarrow\{i\in\mathcal{R}_s\mid \Gamma_i^s=0\}\ \mathrm{sort}\ \mathcal{P}_i^s\uparrow$\tcp*{playback risk}
  $\mathcal{C}_1\leftarrow\{i\in\mathcal{R}_s\mid \Gamma_i^s=1\}\ \mathrm{sort}\ \mathcal{A}_i\downarrow$\tcp*{first audio}
  $\mathcal{C}_2\leftarrow\{i\in\mathcal{R}_s\mid \Gamma_i^s=2\}\ \mathrm{sort}\ \mathcal{U}_i^s\downarrow$\tcp*{U2 efficient}
  $\mathcal{C}\leftarrow \textsc{Concat}(\mathcal{C}_0,\mathcal{C}_1,\mathcal{C}_2)$\tcp*{priority order}
  \ForEach{$i\in\mathcal{C}$}{
    \If{not \textsc{FitsBudget}$(i,\mathcal{B}_s,\mathcal{M}_s)$}{\textbf{break}\;}
    $\mathcal{B}_s\leftarrow \mathcal{B}_s\cup\{i\}$\;
    $\mathcal{M}_s\leftarrow \textsc{UpdateBudget}(\mathcal{M}_s,i)$\;
  }
  \Return $\mathcal{B}_s$\;
}
\caption{\name{} scheduling procedure.}
\label{alg:interactive-scheduling}
\end{algorithm}

This procedure gives realtime requests strict precedence over efficiency-oriented work.
U0 protects playback continuity by serving the request closest to audio drain first, while U1 protects the first-audio path with FCFS-style aging.
U2 is considered only after these classes, where $\mathcal{U}_i^s$ acts as a lightweight ordering heuristic rather than a global optimum or an exact knapsack solution.
The same procedure applies across stages with stage-specific playback-buffer estimates, utility weights, and resource budgets.

\section{Interaction-aware KV Cache Management}
\label{sec:kv}

Section~\ref{sec:challenges} shows that multi-turn Omni realtime interaction turns LLM-stage KV\footnote{We use LLM-stage KV to refer to KV cache maintained by autoregressive model stages, such as the thinker and talker in speech-oriented Omni-LMs.} into long-lived session state.
Reusing LLM-stage KV avoids re-prefilling conversation history and shortens next-turn TTFP.
As HBM capacity is limited, systems~\cite{cachedattention, mooncake} offload idle KV to host DRAM and use least-recently-used (LRU) replacement to keep only a small working set on GPU.
However, deciding which KV to keep on GPU is different in realtime interaction.
Evicting KV that will be reused soon forces a DRAM-to-HBM reload onto the next-turn critical path, while LRU ranks KV by past access time rather than by when the session is likely to speak again.

\begin{figure}[t]
  \centering
  \includegraphics[width=0.90\linewidth]{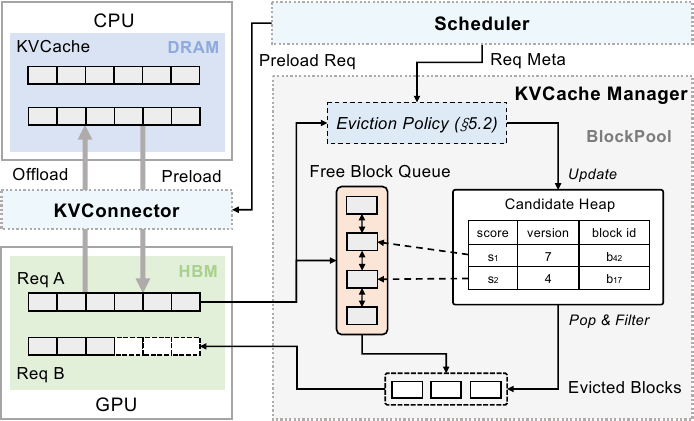}
  \caption{Overview of the \name{}'s KV manager. It manages multi-turn KV residency across HBM and DRAM.}
  \Description{KV manager overview with GPU HBM containing pinned running request KV, evictable multi-turn KV, and free KV space; CPU DRAM containing offloaded multi-turn KV; and arrows for eviction and loading controlled by the KV manager.}
  \label{fig:kv-manager-overview}
\end{figure}

\name{} manages this state with the two paths shown in Figure~\ref{fig:kv-manager-overview}.
For eviction, it uses playback progress to choose idle multi-turn KV by predicted next-use time instead of last-use time.
For loading, it uses speech start and barge-in to preload offloaded KV while the user is still speaking.
In this design, GPU HBM holds pinned running KV, reusable multi-turn KV, and free KV space for active execution, while CPU DRAM stores evicted multi-turn KV.
The scheduler decides which requests run in the current batch, whereas the KV manager decides where reusable cross-turn state resides.
The next two subsections describe the eviction policy (Section~\ref{sec:kv-eviction}) and preload policy (Section~\ref{sec:kv-preload}).

\subsection{Next-use-aware KV Eviction}
\label{sec:kv-eviction}

LRU's weakness appears when playback progress differs across sessions.
A session whose LLM stage just finished may still have a long audio response left to play, so its KV is unlikely to be reused soon unless the user barges in.
Another session may have older KV but be near playback completion, making its next turn more likely.
Thus, last-use order can evict near-future KV blocks while keeping KV blocks whose next use is farther away.

\stitle{Next-use estimate.}
\name{} keeps block-level KV allocation, but ranks eviction candidates by an estimate of when the cached session state will next be needed.
For session $i$, the KV manager estimates:
\begin{equation}
\mathcal{T}_{\mathrm{next},i}
= \mathcal{T}_{\mathrm{play},i}
+ \mathcal{T}_{\mathrm{reply},i}.
\label{eq:kv-next-use}
\end{equation}
Here $\mathcal{T}_{\mathrm{play},i}$ is the remaining playback horizon, and $\mathcal{T}_{\mathrm{reply},i}$ estimates the interval from playback completion to the next completed user input using a per-session moving average when available and a workload-level prior otherwise.

The estimate is used only to order eviction candidates, so it need not be an exact wall-clock prediction.
When playback accounting is unavailable or noisy, \name{} falls back to observable progress counters, using longer unfinished responses as a sign of later reuse.
If the monitor reports speech start or barge-in, \name{} treats the session as immediate reuse and protects its resident KV from normal eviction.

\stitle{Eviction policy.}
When HBM pressure requires freeing KV capacity, the KV manager considers only idle-resident multi-turn KV.
Pinned running KV and sessions marked for immediate reuse are excluded.
The manager orders eligible sessions by decreasing $\mathcal{T}_{\mathrm{next}}$ and scans from the session whose next use is farthest away.
It evicts blocks from that session until enough HBM is released or the session has no evictable blocks left, then continues with the next session if needed.
Thus, \name{} changes the order of eviction targets while keeping allocation and eviction block-level.

Within a selected session, \name{} gives suffix blocks higher eviction priority than prefix blocks. Prefix blocks are shared by more future turns and are more expensive to reconstruct, so keeping them preserves prefix continuity and improves KV reuse. Suffix blocks lie near the tail of the cached conversation state, tend to have a shorter reuse horizon, and therefore impose lower reconstruction cost and smaller impact on subsequent turns when evicted.

This policy preserves the block allocator used by existing serving engines while changing only the eviction order.
As a result, \name{} can free HBM for active requests while reducing short-interval evict-and-reload cycles for sessions close to their next turn.

\subsection{Speech-triggered KV Preloading}
\label{sec:kv-preload}

Next-use-aware eviction reduces unnecessary KV swap-in and swap-out between GPU HBM and host DRAM, but it cannot remove the latency of a necessary reload.
If offloaded KV is loaded only after the next-turn request reaches the LLM stage, the DRAM-to-HBM transfer delays prefill and increases TTFP.
Thus, \name{} starts preload at speech start or barge-in, before the full user input reaches the model.

\stitle{Preload trigger.}
Speech start is an early signal that the next turn is coming, not that the next turn is ready.
After the user starts speaking, the utterance still needs to finish, be encoded, and be routed through the orchestrator.
Barge-in creates a similar window because it interrupts the current response while the user continues speaking a new instruction.
\name{} overlaps this speaking interval with KV transfer, removing part of the DRAM-to-HBM latency from the next-turn critical path.

\stitle{Preload admission.}
When a preload trigger arrives, the KV manager first protects any resident KV of that session from eviction.
If the session KV is offloaded, \name{} treats preload as best-effort background work rather than foreground work.
The manager admits an asynchronous DRAM-to-HBM transfer only when the remaining time before LLM-stage execution is enough to hide the transfer cost under current pressure.
A preload that needs more HBM space may evict later-use idle KV using the policy in Section~\ref{sec:kv-eviction}.
When the admission check fails, \name{} skips the preload and lets the normal LLM-stage path load missing KV.

\name{} bounds background preload so it can't interfere with live work.
Under transient pressure, the admission may wait briefly and re-evaluate, but the time available before LLM-stage execution shrinks during the wait, so preload cannot extend the next-turn deadline.
After a preload completes, \name{} keeps the warmed KV on GPU for a short time so the next turn can reuse it, but caps the total amount of such protected KV.
If the transfer is incomplete, canceled, or evicted before reuse, the next turn falls back to synchronous loading, so preloading affects latency but not correctness.

\section{Implementation}
\label{sec:implementation}

We implement \name{} on top of vLLM-Omni~\cite{vllm-omni}, a stage-oriented extension of vLLM for multimodal serving.
\name{} adds three components to this substrate: an interaction-aware scheduler, a monitor-driven request-state path, and a hierarchical KV manager for multi-turn KV residency.
Together, they are implemented in $\sim$6000 lines of Python.
The implementation preserves the engine's original budget checks and default allocator as fallbacks, so missing metadata or disabled policies reduce to the original serving behavior.

\stitle{Scheduler and monitor integration.}
Each autoregressive stage keeps the original continuous-batching loop, but replaces FCFS admission order with the urgency hierarchy in Section~\ref{sec:scheduling}.
A scheduler mixin reads the monitor's per-session interaction and pressure state at the start of each scheduling round.
It estimates playback buffer, computes the U0/U1/U2 class and U2 utility, and then calls the engine's existing feasibility checks for token budget and available KV blocks.
This keeps the policy change local to request ordering, while memory allocation, preemption, abort handling, and decode execution remain delegated to the underlying engine.

Preload requests use the same scheduling path but are tagged as cancellable background work.
During bursts, the mixin can cancel running preloads or hold new ones before they compete with foreground U0/U1 work.
A canceled preload falls back to synchronous KV loading when the next turn actually runs, preserving the correctness path.

\stitle{Hierarchical KV cache manager.}
The KV manager extends the paged KV block pool and the existing HBM-DRAM offload connector.
Rather than changing the upstream block layout, \name{} stores policy metadata in side tables attached to requests, sessions, and KV blocks.
These side tables track next-use estimates, block position, and preload protection state as turns move between active, idle, offloaded, and preload-protected states.

For eviction, the implementation maintains the indexed candidate heap described in Section~\ref{sec:kv-eviction}, while keeping the original LRU allocator as a fallback on every allocation.
The indexed path can run in logging-only mode or replace LRU only when the metadata needed by the next-use policy is available.
For loading, the entrypoint turns speech-start and barge-in events into best-effort preload requests and uses the offload connector's asynchronous transfer path when admission succeeds.
Cold misses, admission skips, and canceled preloads fall back to the normal foreground load path.

\stitle{Fail-closed operation and instrumentation.}
All \name{} mechanisms are optional at runtime and fail back to the substrate's original behavior.
Missing playback-buffer telemetry disables interaction-aware ordering, while missing U2 utility inputs reduce only U2 ordering to ready-age ordering.
Sparse eviction metadata returns the block pool to default LRU, and failed preload admission returns the next live request to synchronous KV loading.
The prototype records counters for policy fallbacks and preload outcomes, which we use to validate in evaluation that the scheduler and KV manager exercise the intended design paths.

\section{Experimental Evaluation}
\label{sec:evaluation}

\name{} is designed to improve the serving performance of Omni realtime interaction workload.
We evaluate whether our interaction-aware policies improve user-facing latency and serving efficiency over existing serving systems.
Our evaluation focuses on the following questions:
\squishlist
\item How does \name{} compare with baselines under realtime interaction workloads, notably with user barge-in?
\item How much does interaction-aware scheduling improve system performance under the playback-continuity SLO?
\item How much does interaction-aware KV management reduce reload overhead for multi-turn sessions?
\squishend

\begin{figure}[t]
  \centering
  
  \includegraphics[width=0.95\linewidth]{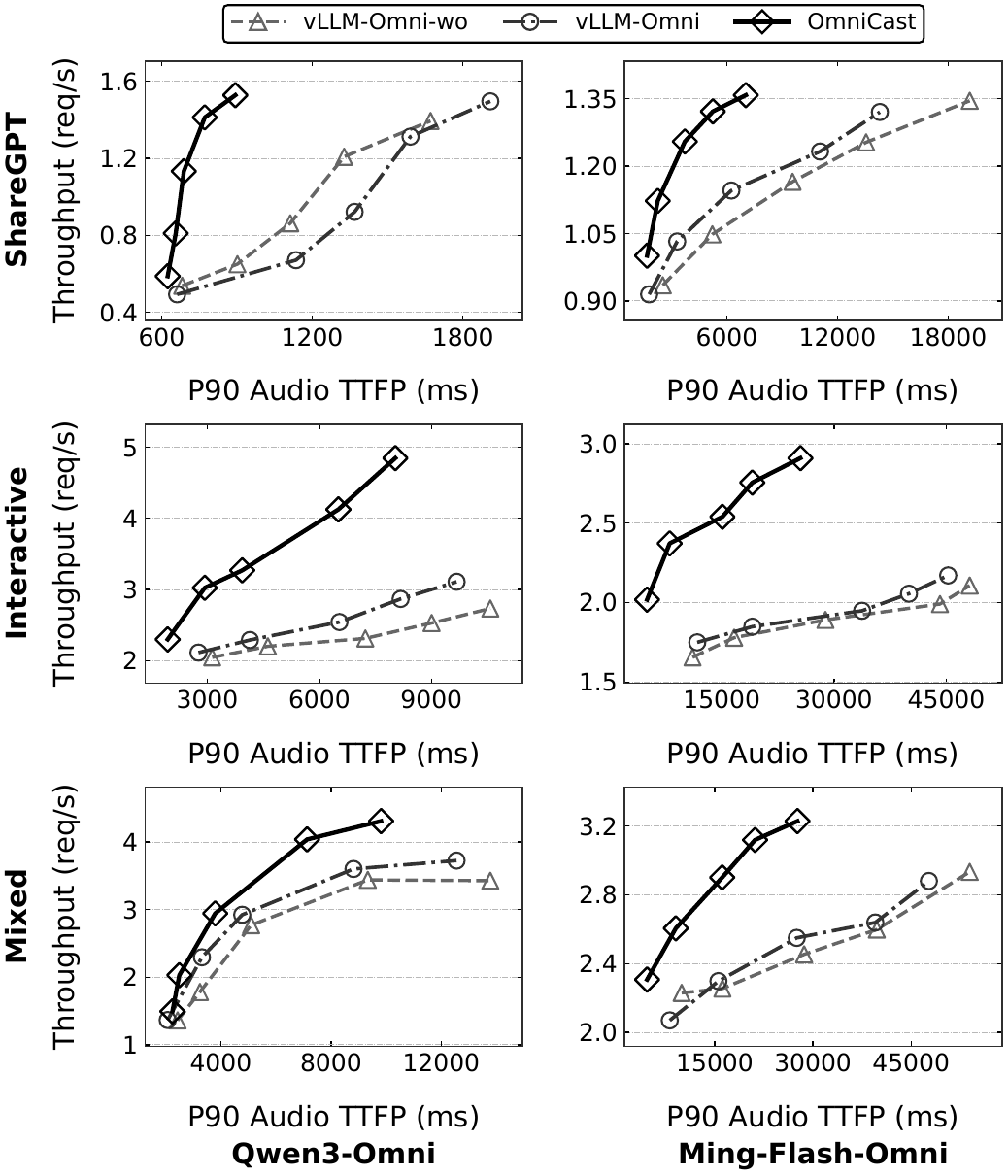}
  \caption{End-to-end throughput-latency frontier across two Omni-LMs and three workloads. Each curve connects results over concurrency-pressure values $c\in\{2,4,8,12,16\}$; higher and further-left points are better.}
  \label{fig:e2e-qwen-workloads}
\end{figure}

\subsection{Experiment Settings}
\label{sec:setup}

\stitlestart{Datasets.}
We evaluate \name{} with three complementary data sources that cover single-turn speech, multi-turn voice interaction, and mixed modality Omni requests.
\squishlist
  \item \textit{ShareGPT conversations.} The single-turn workload is constructed
  from both short and long conversations to stress first-token latency under different context lengths. We use the ShareGPT Chinese-English 90K corpus \cite{sharegpt90k} as the source of realistic conversational prompts.

  \item \textit{Interactive traces.} The multi-turn workload is built from
  retained interaction traces ~\cite{multi-turn-trace}. Each record has a session ID, a request timestamp, query and response token lengths, and a turn index. This structure preserves repeated user turns and model speech replies within the same session.

  \item \textit{StreamingBench and mixed Omni data.} The video and multimodal portion uses StreamingBench \cite{streamingbench} as the source of video questions and media assets. We combine these video events with retained interactive voice sessions to form mixed text, speech, and video workloads.
\squishend

\stitle{Workloads.}
We generate online request arrivals using both synthetic and trace-driven
patterns. For the synthetic setting, requests arrive according to a
\textit{Poisson} distribution, and we evaluate a range of offered loads to measure the highest request rate each system can sustain under the realtime SLO. For trace-driven setting, we use \textit{BurstGPT}~\cite{Burstgpt} arrivals to capture short-term load spikes in interactive services.

We model user barge-in as a client-side policy. Each request independently triggers a barge-in event with Bernoulli probability $p_{\mathrm{bi}}$; unless a barge-in experiment is explicitly reported, we set $p_{\mathrm{bi}}=0$. Barge-in sensitivity experiments use $p_{\mathrm{bi}}\in\{0.0,0.3,0.7,1.0\}$ to cover different barge-in regimes. For requests with barge-in, the cut time is anchored at audio TTFP and then sampled from the dataset-derived distribution of output audio durations. 

\stitle{Baselines.}
We compare \name{} against two baselines. The first is vLLM-Omni~\cite{vllm-omni} without KV offloading, denoted as \textit{vLLM-Omni-wo}. It serves Omni-LM stages with the default scheduling and memory management policies. The second is vLLM-Omni with vLLM-style KV offloading enabled for multi-turn sessions, which offloads KV cache to DRAM and uses LRU to maintain the GPU HBM working set. Unless otherwise specified, KV offloading is enabled by default. 

\stitle{Models and deployment.}
We use two representative speech-oriented Omni-LMs: Qwen3-Omni~\cite{Qwen3-Omni} and Ming-Flash-Omni 2.0~\cite{Ming-Flash-Omni}. We follow the official vLLM-Omni deployment configurations and scale them to an 8-GPU server. Qwen3-Omni uses a three-stage pipeline\footnote{By default, Qwen3-Omni colocates the encoder with the thinker. Its vocoder is a lightweight CNN module and is colocated with the talker.}, with DP=4 for the thinker, DP=4 for the talker. Ming-Flash-Omni 2.0 uses a two-stage pipeline with a TP=2, DP=2 thinker and a DP=4 talker stage.

All compared systems share the same hardware allocation, model weights, and serving configurations.

\stitle{Metrics.}
We evaluate \name{} with metrics that reflect the user-facing goals of Omni realtime interaction: fast first audio, smooth playback, and serving throuhgput.

\squishlist
  \item \textit{Audio time-to-first-packet (TTFP)} is the elapsed time from the end of the replayed user turn, as observed by the benchmark client, to the first decodable audio fragment delivered by the server. By default we report P90 TTFP.

  \item \textit{Real-time factor (RTF)} is the ratio between audio generation time and the audio duration. RTF $<1$ indicate that the system generates audio faster than real-time playback.

  \item \textit{Playback continuity} measures whether the streamed audio plays smoothly after the first chunk arrives. A request is continuous if playback gaps stay below 100ms, the default threshold in vLLM-Omni benchmark. Requests with barge-in are excluded unless stated otherwise.

  \item We report throughput as \textit{requests per second (RPS)} over the steady-state window. \textit{Useful RPS} is the highest offered load at which the system meets the continuity SLOs.
\squishend

\stitle{Testbed.}
Experiments were conducted on a single H200 server with eight NVIDIA H200 GPUs.
The host contains two Intel Xeon Platinum 8480C sockets, 56 physical cores per
socket, and 2.0 TiB of DRAM. Our implementation is developed based on vLLM
0.20.0 and vLLM-Omni 0.20.1.

\subsection{Main Results}
\label{sec:e2e}

\stitlestart{Throughput-latency frontier.}
Figure~\ref{fig:e2e-qwen-workloads} reports the end-to-end throughput--latency frontier as the c-bound concurrency of online sessions changes. The top row uses Qwen3-Omni and the bottom row uses Ming-Flash-Omni 2.0; within each row, the panels correspond to ShareGPT, interactive, and mixed workloads. The x-axis is P90 audio TTFP and the y-axis is completed request throughput, so better systems move toward the upper-left region.

Across the two Omni-LMs and three workloads, \name{} consistently improves the frontier over both vLLM-Omni baselines. On ShareGPT, where the workload is less dominated by multi-turn KV reuse, \name{} achieves comparable or higher peak throughput while reducing high-concurrency P90 audio TTFP by about $2\times$ on both models.

On interactive traces, repeated turns increase KV pressure and scheduling contention, making the gap larger. For Qwen3-Omni, \name{} improves peak throughput by $56$--$78\%$ over the baselines while also reducing P90 audio TTFP at the same concurrency; at moderate concurrency, it further improves throughput by $28.5\%$ and lowers P90 audio TTFP by $39.8\%$ over the offloading baseline. Ming-Flash-Omni 2.0 shows the same upward-left shift, indicating that the benefit is not specific to one Omni-LM.

On the mixed workload, \name{} remains effective under heterogeneous stage demand. At the largest c-bound, it improves throughput by $12$--$16\%$ over the best baseline and substantially reduces P90 audio TTFP, with the reduction reaching about $42\%$ on Ming-Flash-Omni 2.0. Overall, simple KV offloading makes longer sessions feasible, but without interaction-aware scheduling and KV management, the saved memory does not translate into the same first-audio responsiveness.

\begin{figure}[t]
  \centering
  \includegraphics[width=0.90\linewidth]{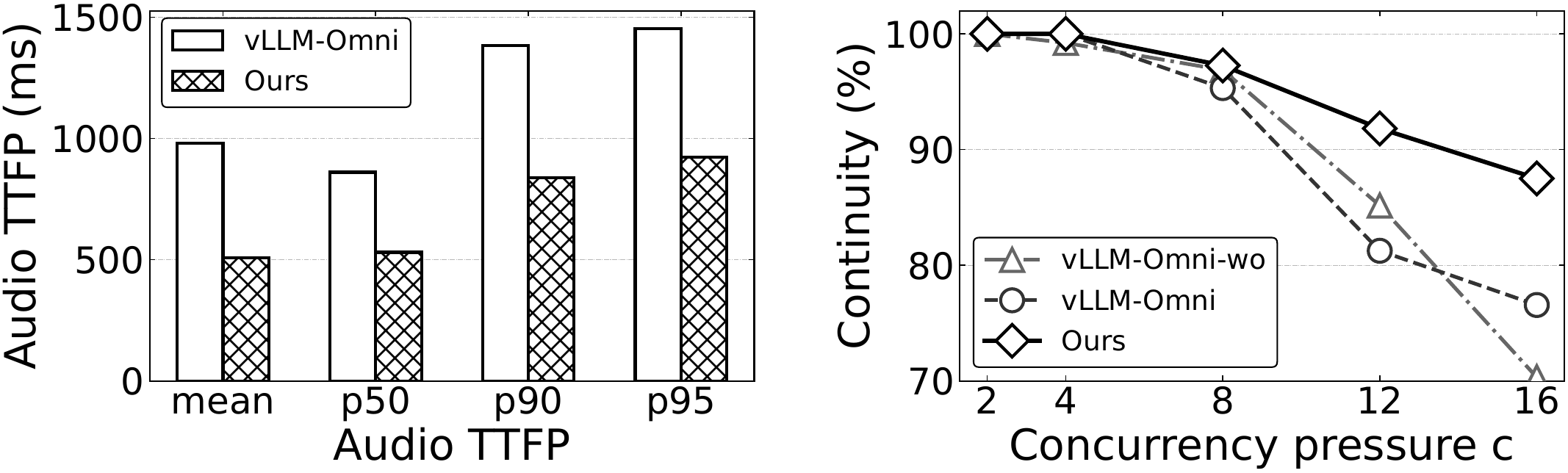}
  \caption{Interactive playback continuity and generated-token waste under concurrency and barge-in pressure.}
  \label{fig:latencytail-and-waste}
\end{figure}

\begin{figure}[t]
  \centering
  \includegraphics[width=0.90\linewidth]{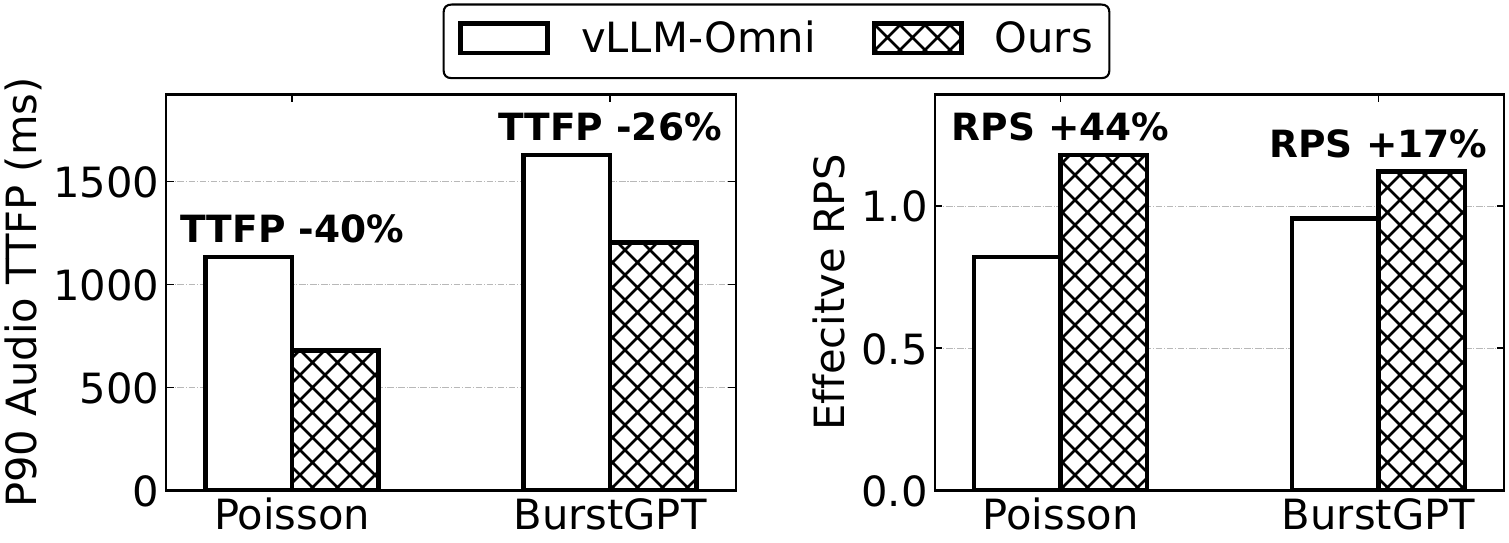}
  \caption{Effect of interaction-aware scheduling under Poisson and BurstGPT arrivals using Qwen3-Omni audio serving.}
  \Description{Two grouped-bar panels compare vLLM-Omni with KV offloading and LiveServe. The latency panel reports lower P90 audio TTFP for LiveServe, and the throughput panel reports higher effective RPS for LiveServe under both Poisson and BurstGPT arrivals.}
  \label{fig:interactive-distribution-benefit}
\end{figure}

\stitle{Tail-latency distribution at fixed concurrency.}
The left panel of Figure~\ref{fig:latencytail-and-waste} drills into the Qwen3-Omni ShareGPT audio workload without configured barge-in at $c=8$. \name{} lowers the whole visible latency distribution: the median decreases from $0.86$~s to $0.53$~s, while P90 and P95 fall from $1.38$~s and $1.45$~s to $0.84$~s and $0.92$~s. This shows that the scheduler compresses both typical and tail first-audio delays under the same concurrency pressure.

\stitle{Playback continuity.}
The right panel of Figure~\ref{fig:latencytail-and-waste} evaluates whether the lower latency comes at the cost of streamed-audio continuity. Under heavier pressure, \name{} degrades more gracefully: at $c=12$, continuity remains $91.8\%$, compared with $81.3\%$ for vLLM-Omni with offloading; at $c=16$, \name{} keeps $87.5\%$ continuity, compared with $76.6\%$ and $70.3\%$ for the two baselines. Thus, the U0/U1/U2 scheduler preserves enough playback buffer for continuity while reducing work on well-buffered U2 sessions.

\stitle{Arrival distribution.}
Figure~\ref{fig:interactive-distribution-benefit} compares \name{} with the KV-offloading baseline under Poisson and trace-driven BurstGPT arrivals. This experiment uses an audio ShareGPT-style workload with $c=8$, $32$ requests, request rate $4$~RPS for the Poisson case, and a BurstGPT manifest with the same peak request rate. Under Poisson arrivals, \name{} reduces P90 audio TTFP from $1.13$~s to $0.68$~s and raises effective throughput from $0.82$ to $1.18$~RPS. Under BurstGPT arrivals, where short-term bursts leave less scheduling slack, \name{} still lowers P90 audio TTFP from $1.63$~s to $1.20$~s and improves effective throughput from $0.96$ to $1.12$~RPS. This shows the scheduling policy remains robust even under bursty arrivals.

\begin{figure}[t]
  \centering
  \includegraphics[width=0.90\linewidth]{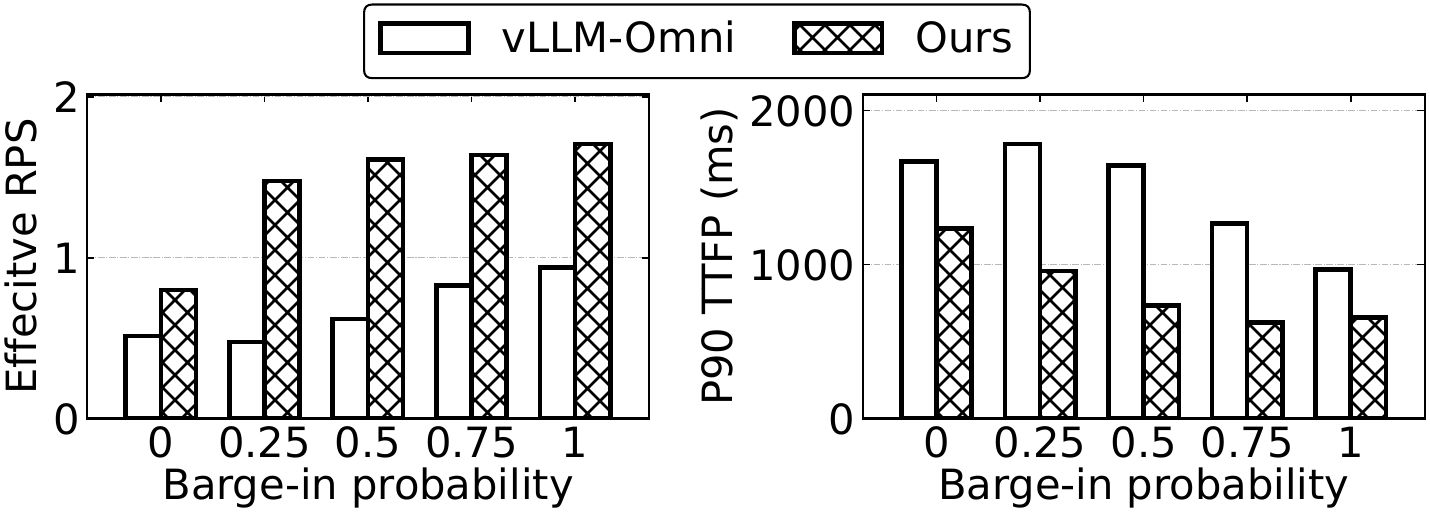
  }
  \caption{Sensitivity to configured barge-in probability on the ShareGPT audio workload.}
  \label{fig:sharegpt-interrupt-rps-ttfp}
\end{figure}

\stitle{Sensitivity to user barge-in.}
Figure~\ref{fig:sharegpt-interrupt-rps-ttfp} reports sensitivity to configured ShareGPT barge-in probability. The experiment uses Qwen3-Omni audio serving with $c=8$. On effective RPS, \name{} outperforms the offloading baseline across the full range: at $p_{\mathrm{bi}}=0.5$ and $p_{\mathrm{bi}}=0.75$, throughput improves by $2.6\times$ and $2.0\times$, respectively. This experiment shows the corresponding P90 audio TTFP, where \name{} is also consistently lower, cutting latency by more than half at the same two barge-in probabilities. The improvement comes from the playback-buffer-aware urgency hierarchy: U0/U1 requests stay ahead of well-buffered U2 sessions, while the U2 barge-in exposure term limits discardable audio work before an abort arrives.

\begin{figure}[t]
  \centering
  \includegraphics[width=0.9\linewidth]{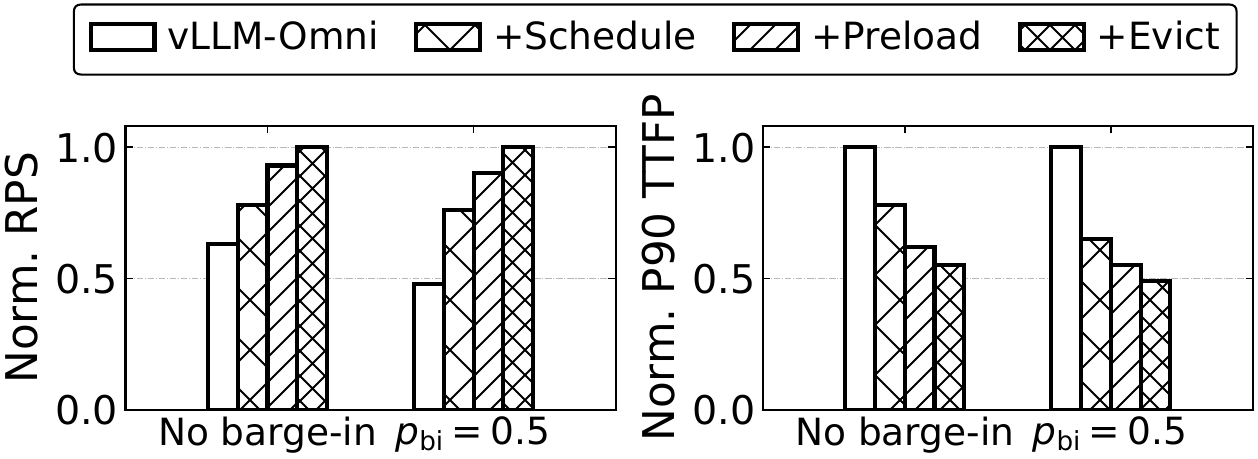}
  \caption{Component ablation on the interactive workload.}
  \Description{Grouped bar chart showing incremental gains from the interaction-aware scheduler, KV-cache preload, and interaction-aware KV-cache eviction.}
  \label{fig:micro-ablation}
\end{figure}

\subsection{Analysis}
\label{sec:analysis}

We next run controlled experiments to study where \name{}'s gains come from and how robust they are across workload conditions. Unless otherwise stated, this analysis uses Qwen3-Omni in audio mode.

\begin{figure}[t]
  \centering
  \includegraphics[width=0.9\linewidth]{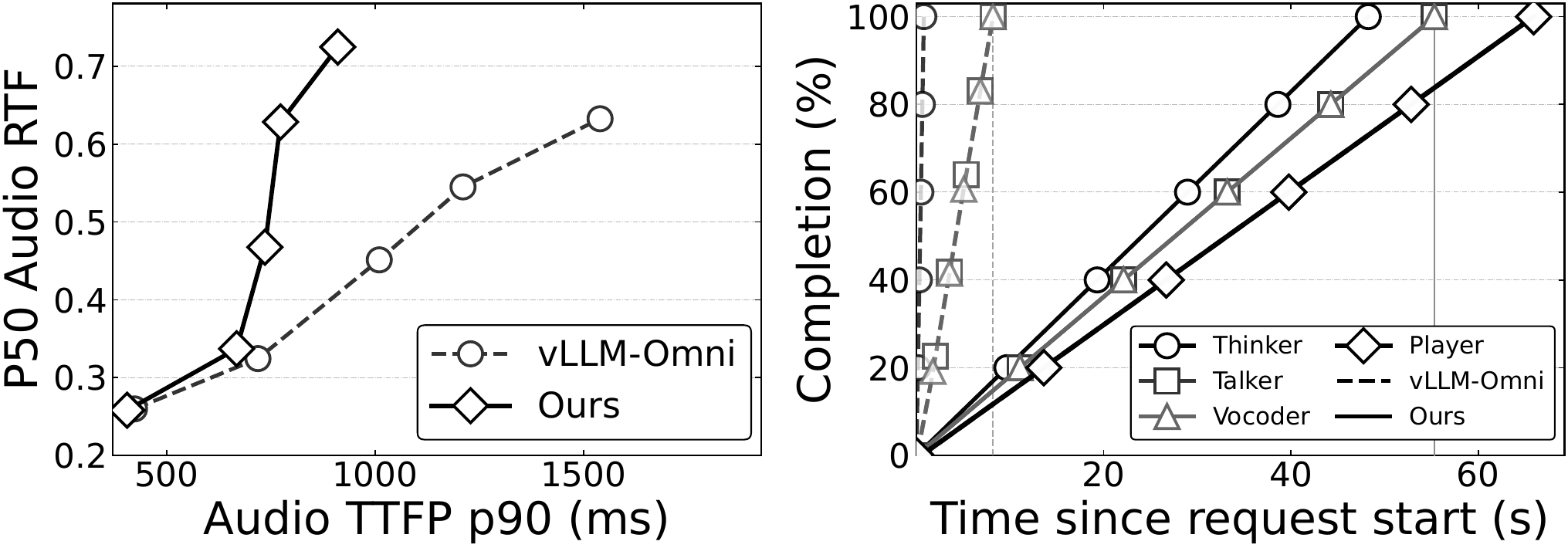}
  \caption{Audio generation pacing. The left panel varies concurrency on a ShareGPT audio workload; the right panel illustrates generation and playback completion over time.}
  \Description{Two panels shown side by side. The left panel is a line plot comparing the baseline without back-pressure and LiveServe with playback-buffer-aware scheduling as concurrency increases from 1 to 8; LiveServe keeps audio RTF below 1 while reducing P90 audio TTFP. The right panel shows long-horizon generation and playback progress.}
  \label{fig:rtf-ttfp-frontier}
\end{figure}

\stitle{Component ablation.}
Figure~\ref{fig:micro-ablation} adds \name{}'s components one by one using Qwen3-Omni under two settings: without barge-in and with barge-in probability $p_{\mathrm{bi}}=0.5$. Without barge-in, the full system reduces P90 TTFP by $29.8\%$ and improves RPS by $8.8\%$. With barge-in enabled, the gains increase to $39.8\%$ lower P90 TTFP and $28.5\%$ higher RPS. The staged improvements show that scheduling, preload, and eviction are complementary rather than redundant.

\stitle{RTF-latency tradeoff.}
Figure~\ref{fig:rtf-ttfp-frontier} shows why \name{} does not simply generate speech as fast as possible. In the ShareGPT audio workload with $p_{\mathrm{bi}}=0.5$, both systems keep P90 RTF below $1$, so playback can remain faster than real time. However, as concurrency grows, \name{} converts work on well-buffered U2 sessions into lower first-audio latency. At $c=8$, P90 audio TTFP drops from $1.54$~s to $0.91$~s, while median RTF remains below real time. The long-horizon example explains the mechanism: the baseline finishes model-side generation in about $8.2$~s while the player consumes the response over about $65.9$~s, accumulating a large stage-aware playback buffer that can be discarded after barge-in. \name{} stretches generation to about $55.3$~s, closer to playback progress, preserving enough buffer for continuity while freeing decode capacity for U1 first-audio requests.

\begin{figure}[t]
  \centering
  \includegraphics[width=0.9\linewidth]{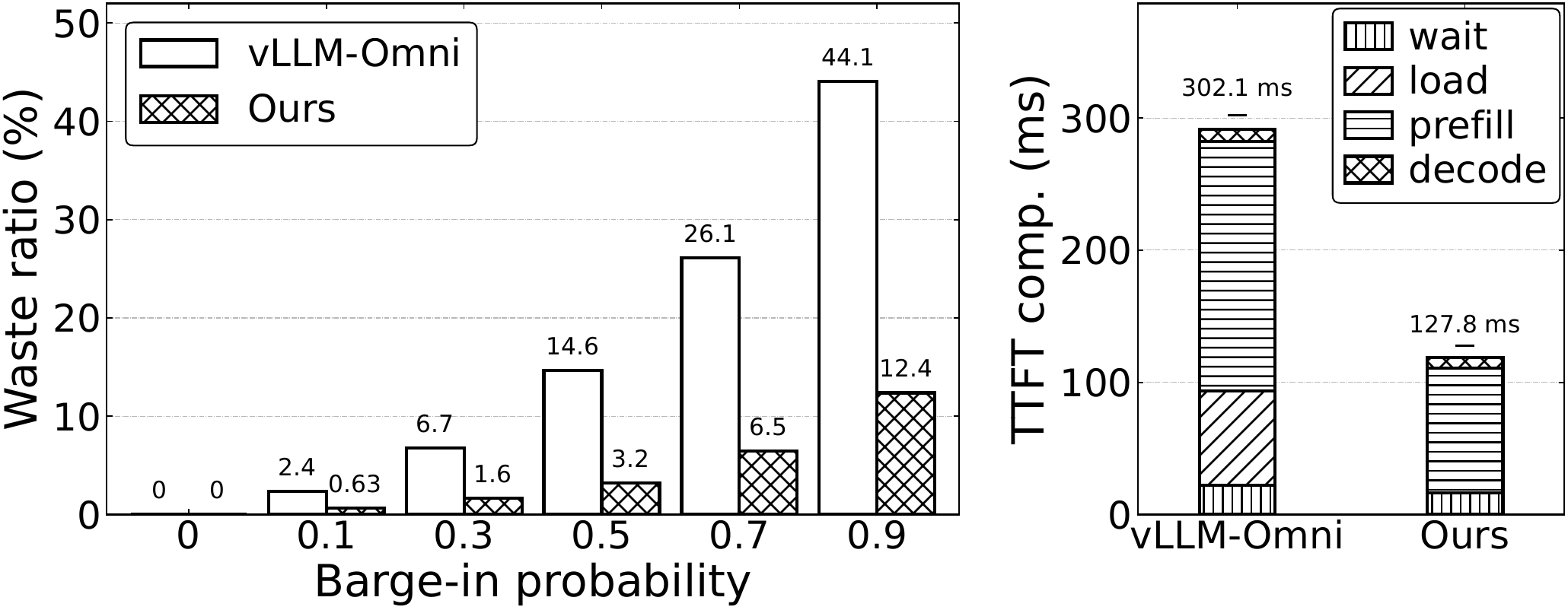}
  \caption{Impact of barge-in and reload pressure. Left: wasted-token ratio under different barge-in probabilities. Right: latency breakdown of a reload-pressure target request.}
  \Description{Two latency breakdown plots shown side by side: one reports segment latencies under KV reload pressure, and the other decomposes text time-to-first-token.}
  \label{fig:latbd-breakdowns}
\end{figure}

\stitle{Barge-in token waste.}
The left panel of Figure~\ref{fig:latbd-breakdowns} profiles generated-but-unheard production tokens under different configured barge-in probabilities. With no barge-in, both systems waste no generated tokens. As the barge-in probability increases, the vLLM-Omni waste ratio rises to $44.06\%$, because generation can run far ahead of playback until an abort arrives. \name{}'s U2 barge-in exposure term limits this discardable buffered work, reducing the waste ratio to at most $12.38\%$ and eliminating about $72\%$--$78\%$ of wasted generated tokens across barge-in settings.

\stitle{First-token critical path under reload pressure.}
The right panel of Figure~\ref{fig:latbd-breakdowns} shows a multi-turn request from an interactive workload, where interaction-aware KV management removes reload work from the user-visible path. The offloading baseline spends $71.0$~ms on on-path KV reload and reaches $302.1$~ms text TTFP. With a warm prefetch hit, \name{} eliminates the reload segment and reduces text TTFP to $127.8$~ms, a $57.7\%$ reduction, by moving bulk HBM--DRAM transfer off the next-turn critical path.

Together, these analyses show that \name{}'s gains come from complementary mechanisms. Playback-aware scheduling protects U0/U1 requests and limits discardable U2 work, improving effective throughput and first-audio latency under barge-in. Interaction-aware KV management ensures that when a session returns for its next turn, the required state is already resident or prefetched.

\subsection{Microbenchmarks}
\label{sec:micro}

\begin{figure}[t]
  \centering
  \includegraphics[width=0.90\linewidth]{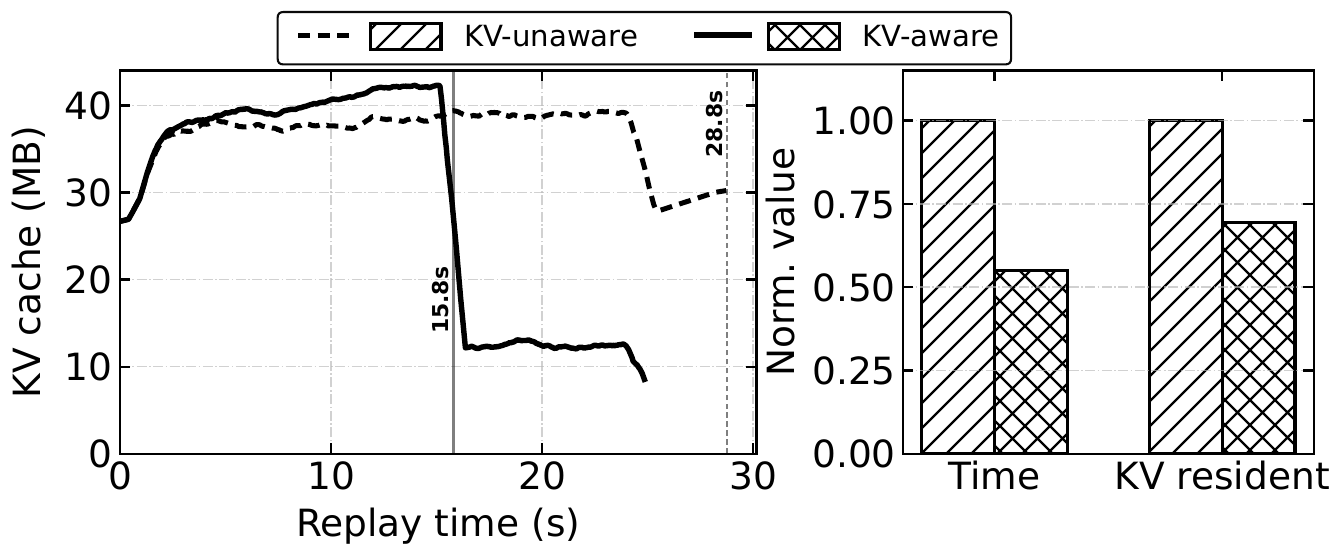}
  \caption{Effect of KV-aware deferral on thinker KV residency. Left: thinker GPU KV timeline. Right: normalized completion time and resident-KV footprint.}
  \Description{Two-panel plot comparing KV-unaware and KV-aware deferral policies. The left panel shows thinker GPU KV-cache occupancy over replay time, and the right panel shows normalized time and KV-resident metrics.}
  \label{fig:micro-thinker-kv-occupancy}
\end{figure}

\stitlestart{KV residency timeline.}
Figure~\ref{fig:micro-thinker-kv-occupancy} profiles KV-aware U2 scheduling on the Qwen3-Omni interactive workload. Under KV pressure, KV-aware ordering favors resident long-context requests, allowing them to finish and release HBM earlier than KV-unaware ordering. The right panel summarizes this effect with a shorter replay and lower normalized resident-KV footprint, showing why efficiency-class scheduling should account for memory residency.

\begin{figure}[t]
  \centering
  \begin{subfigure}[t]{0.49\linewidth}
    \centering
    \includegraphics[width=\linewidth]{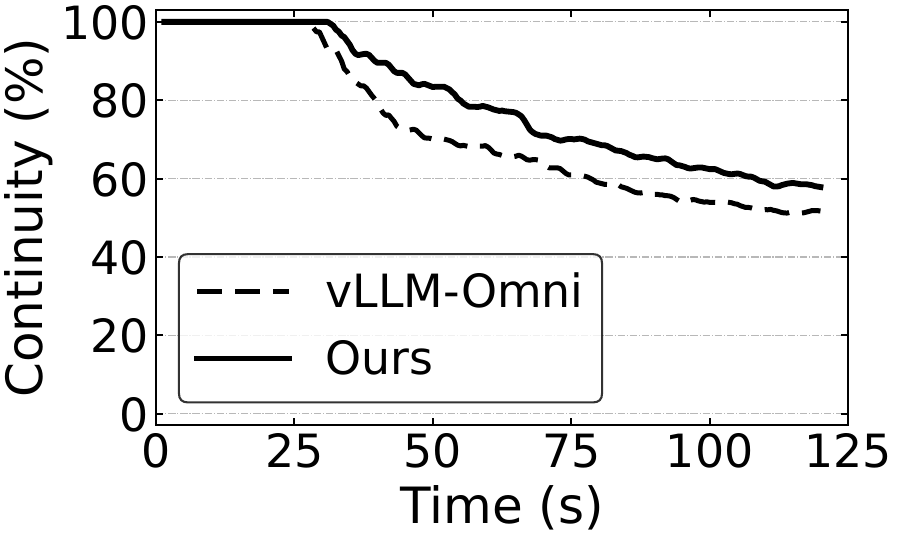}
    \caption{No barge-in}
    \label{fig:micro-continuity-timeline-nointerrupt}
  \end{subfigure}\hfill
  \begin{subfigure}[t]{0.49\linewidth}
    \centering
    \includegraphics[width=\linewidth]{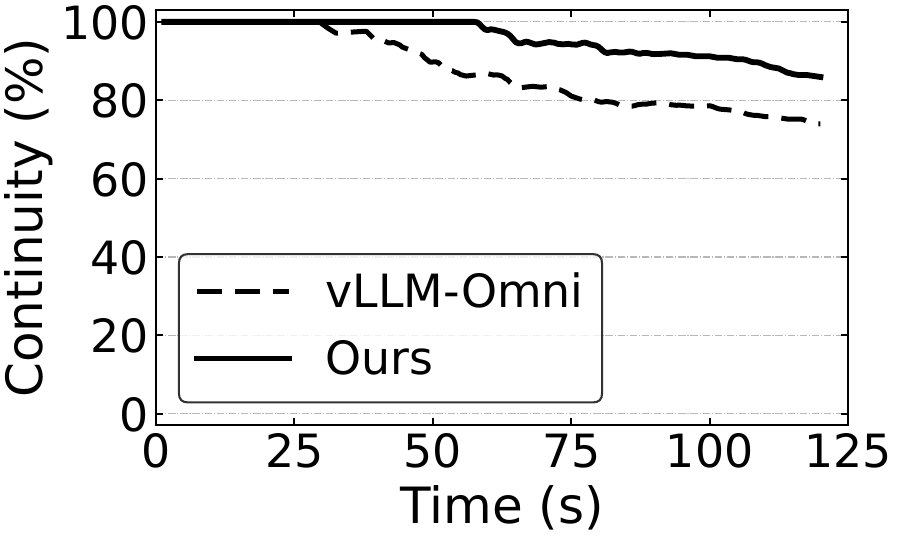}
    \caption{$p_{\mathrm{bi}}=0.5$}
    \label{fig:micro-continuity-timeline-bargein}
  \end{subfigure}
  \caption{Playback-continuity timeline. The right panel enables barge-in with triggers anchored after TTFP.}
  \Description{Two continuity timeline plots shown side by side, comparing vLLM-Omni and LiveServe over time under no-barge-in and barge-in settings.}
  \label{fig:micro-continuity-timeline}
\end{figure}

\stitle{Playback-continuity timeline.}
Figure~\ref{fig:micro-continuity-timeline} stress-tests playback continuity over time on ShareGPT with BurstGPT arrivals ($c=32$, $12$~RPS). Without barge-in, both systems start at full continuity, but \name{} ends higher than vLLM-Omni. With barge-in at $p_{\mathrm{bi}}=0.5$, \name{} keeps a clearer advantage by limiting U2 barge-in exposure and avoiding obsolete in-flight audio work.

\begin{table}[t]
\centering
\caption{Effect of the eviction index optimization.}
\label{tab:eviction-index-overhead}
  \fontsize{9}{11}\selectfont
\setlength{\tabcolsep}{3.5pt}
\begin{tabular}{lcccc}
\toprule
System & Avg. OH & P90 OH & Effective  & E2E P90 \\
       & (ms) & (ms) & RPS & (ms) \\
\midrule
\name{}   & 0.093 & 0.222 & 2.625 & 3069 \\
w/o index   & 5.311 & 8.270 & 1.970 & 3381 \\
\bottomrule
\end{tabular}
\end{table}

\stitle{Eviction-index overhead.}
Table~\ref{tab:eviction-index-overhead} measures eviction overhead in the interactive multi-turn dialogue scenario without barge-in. Compared with tail scanning, \name{}'s heap-based eviction index reduces average overhead from $5.31$~ms to $0.093$~ms and P90 overhead from $8.27$~ms to $0.222$~ms, while improving effective QPS from $1.97$ to $2.63$. This shows that indexed eviction avoids making next-use-aware eviction a scheduler bottleneck.

\stitle{Takeaway.}
The microbenchmarks validate the key design assumptions behind the end-to-end gains. For memory management, the goal is not to minimize KV residency at all times, but to keep the right session KV resident before its next turn. For audio scheduling, the scheduler should preserve enough playback buffer for continuity without accumulating excessive discardable audio. Together, next-use-aware eviction, and speech-triggered preload improve both memory stability and user-visible latency in multi-turn realtime interaction.

\section{Related Work}
\label{sec:related}

\stitle{Efficient LLM serving.}
Many systems optimize LLM serving from different parts of the inference pipeline~\cite{orca, vllm, sglang, sarathi, distserve, tokenflow, andes}.
vLLM~\cite{vllm} and SGLang~\cite{sglang} are the most popular open-sourced repositories, and they cooperated with system optimizations such as continuous batching~\cite{orca}, prefill/decode disaggregation~\cite{distserve, splitwise}, and chunked prefill~\cite{sarathi, deepspeed-fastgen} to improve the overall system performance of serving LLMs.
Some systems~\cite{fastserve, llumnix, fairness} propose scheduling algorithms to handle dynamic workloads.
Besides, Andes~\cite{andes} and TokenFlow~\cite{tokenflow} observe that LLMs often generate text faster than users can consume it, so continuously decoding can waste GPU time without improving perceived quality. They schedule by QoE and token-buffer status, pausing well-buffered requests to serve stall-prone ones.
In contrast, \name{} focuses on a more complex multi-stage Omni-LM pipeline, where scheduling follows audio playback progress to avoid barge-in waste from each LLM stage running too far ahead.

\stitle{Multi-turn KV management.}
Multi-turn interaction turns KV cache from a per-request workspace into reusable session state, motivating systems that preserve conversation history across turns instead of recomputing the full prompt~\cite{cachedattention, flashgen, mooncake, lmcache, pensieve, hcache, cachegen, multi-turn-trace}.
CacheAttention~\cite{cachedattention} and FlashGen~\cite{flashgen} use multi-tier caching and request scheduling to accelerate multi-turn serving.
Mooncake~\cite{mooncake} builds a KV-centric disaggregated architecture that pools cache storage across inference nodes.
LMCache~\cite{lmcache} provides a KV cache layer spanning GPU, CPU, disk, and remote storage.
Pensieve~\cite{pensieve} manages GPU-CPU KV residency for conversation state, evicting cached chunks based on inactivity and recomputation cost while supporting non-contiguous cached context.
These systems optimize KV reuse, capacity, and transfer cost, while \name{} focuses on when session KV should reside in GPU memory under realtime interaction, where audio playback and barge-in determine the next-turn latency path.

\stitle{Serving systems for multimodal models.}
Multimodal serving spans both understanding models that consume non-text inputs and produce text, and generation models that also synthesize speech, images, or video.
Early systems mainly focused on multimodal understanding workloads.
EPD disaggregation~\cite{epd}, EPD-Serve~\cite{epdserve}, HydraInfer~\cite{hydrainfer}, and SpaceServe~\cite{spaceserve} disaggregate multimodal encoding, LLM prefill, and LLM decode into different workers to reduce encoder-induced interference.
ModServe~\cite{modserve}, ElasticMM~\cite{elasticmm}, and TCM-Serve~\cite{tcm-serve} allocate resources according to modality- and stage-level workload heterogeneity.

Generation-oriented Omni serving further complicates the problem because heterogeneous stages also appear on the output side. Existing systems either expose Omni inference as stage-oriented runtimes~\cite{vllm-omni,sglang-omni}, optimize generic any-to-any computation graphs~\cite{cornserve}, or target streaming speech models with speech-aware scheduling~\cite{voxserve}. In contrast, \name{} targets realtime Omni interaction with interaction-aware scheduling and KV management for playback, barge-in, first-audio latency, and multi-turn sessions.

\section{Conclusion}
\label{sec:conclusion}

Existing stage-oriented Omni serving systems ignore live interaction signals, causing over-generation and poorly timed multi-turn KV management.
We present \name{}, which tracks playback progress and speech activity, and applies interaction-aware scheduling together with next-use-aware KV eviction and preloading.
Across Omni-LM models and realtime workloads, \name{} lowers P90 audio TTFP up to $2.21\times$, while improving completed-request throughput up to $1.56\times$, and cuts wasted generated tokens by $72$-$78\%$ under barge-in. \name{} shows that making Omni serving aware of live interaction state can substantially improve both system efficiency and user-perceived responsiveness.


\bibliographystyle{ACM-Reference-Format}
\bibliography{references}

\end{document}